\documentclass[12pt,a4paper,twoside]{article}
\usepackage{Abbreviations}\usepackage{graphicx}
\usepackage{amsmath}\usepackage{bbm,bm,amsfonts,amssymb}

\begin{document}
\title{Entropy in some simple one-dimensional configurations}
\author{M. Bordag\thanks{bordag@uni-leipzig.de}\\
\small Universit{\"a}t Leipzig}
\date{26.7.2018}

\maketitle

\begin{abstract}
Continuing the discussion on negative entropy in Casimir-effect like configurations, we consider two simple one-dimensional examples. One is the s-wave contribution to a plasma sphere and the other a single delta function potential. Some information on generic background potential is gained applying Levinson's theorem. For the first example we find negative entropy. The one-dimensional examples are especially interesting as these do not require the subtraction of contributions growing with temperature faster than the classical limit.
\end{abstract}\thispagestyle{empty}

\section{\label{T1}Introduction}
Negative entropy was observed recently in some simple single-object configurations.
In \cite{Text18-1} it was found for the \elm field in the presence of a plasma sphere and in \cite{Text18-2} it was found for flat sheets. These configurations are interesting as providing single bodies in opposite to Casimir effect like configurations where one considered only the separation dependent part of the free energy and of the entropy.  In those configurations negative values for the entropy were reported repeatedly starting with \cite{geye05-72-022111}. However, these constitute only a part of the complete entropy leaving room for compensating positive contributions.

In the present paper we investigate some simple one-dimensional configurations, where the calculations take only a few lines and results are quite explicit. Special interest in one-dimensional configurations comes from the absence of contributions growing faster than the classic limit for $T\to\infty$. In case such contributions are present as happens in higher dimensions, these must be subtracted as explained in the above mentioned papers (see also \cite{BKMM}, Chapt. 5). This subtraction procedure raises questions which are not fully answered yet. If we find negative entropy in a one-dimensional system, where no subtraction is necessary, we can look on the results of \cite{Text18-1} and \cite{Text18-2} with more conviction.

Throughout the paper we use units with $k_{\rm B}=\hbar=c=1$.

\section{\label{T2}Basic formulas}
In this section we introduce in short the basic formulas which we use.  All these formulas can be found in standard literature.

We consider a field $\Phi$ in $(1+1)$-dimensional space-time obeying the equation
\eq{1}{ \left(-\pa_x^2+V(x)\right)\Phi(x)=\om^2\Phi(x)
}
after Fourier transform in time. Here $\om$ is the frequency and $V(x)$ is a background potential. Examples to be considered are
\ben    \item {\it plasma point}
\eq{2}{ V(x)&=\Om\,\delta(x-R),\ \ x\in [0,\infty),
\nn\\       \Phi(0) &=0
}
This is a problem on the half axis with Dirichlet boundary condition at $x=0$. An equivalent understanding is an infinite potential $V(x)$ on the negative half axis.
We call this configuration 'plasma point' since it is the s-wave contribution to the plasma sphere considered in \cite{Text18-1}. According to this origin we call $\Om$ {\it plasma frequency} and restrict it to positive values, $\Om>0$. In that case there are no bound states.

\item  {\it  delta function potential}\\
\eq{3}{ V(x) &=2\al\,\delta(x), \ \ x\in (-\infty,\infty)
}
This potential is a delta function, $\al$ is its strength and it is well known from class room examples in quantum mechanics. For $\al>0$ there are no bound states and for $\al<0$ there is one bound state with binding energy $\kappa=-\al$.
\een

The temperature dependent part of the free energy is given as usual by
\eq{4}{  \Delta_T{\cal F} &= T\sum_n \ln\left(1-\exp\left({-\frac{-\kappa_n+\mu}{T}}\right)\right)
\nn\\  &~~~    +\frac{T}{\pi}\int_0^\infty d\om\, \ln\left(1-\exp\left(-\frac{\om+\mu}{T}\right)\right)\frac{\pa}{\pa\om}\delta(\om),
}
where $\delta(\om)$ is the phase shift of the scattering setup connected with eq. \Ref{1}. The phase shift is by
\eq{5}{\delta(\om)=\frac{1}{2i}\ln\left(\frac{t(\om)}{t(\om)^*}\right)
}
related to the transmission coefficient $t(\om)$ for a problem on the whole axis. For the half axis see below in Sect. \ref{T3}.  The derivative of the phase shift is the density of states. In \Ref{4}, the first term is absent in case there are no bound states. Otherwise, the $\kappa_n$ are the binding energies and at once the locations of the poles of the transmission coefficient on the imaginary frequency axis at $\om=i\kappa_n$. In \Ref{4}, $\mu$ is the chemical potential and it must be larger that the lowest bound state energy, i.e., $\mu>\kappa_n$ must hold for all $n$.

Equivalent to \Ref{4} is the Matsubara representation,
\eq{6}{{\cal F}&= T\sum_{l=-\infty}^{\infty} \ln\left(t(-\xi_l)\right),\ \
            \ \ (\xi_l=2\pi T l),
}
which includes also the vacuum energy. Since the vacuum energy has a ultraviolet divergence, this representation is less convenient for calculating the entropy.

The entropy $S$ is given by the thermodynamic relation
\eq{7}{ S &=-\frac{d}{d\,T}{\cal F}
}
and using \Ref{4} it can be represented in the form
\eq{8}{ S &= \sum_n g\left(\frac{-\kappa_n+\mu}{T}\right)
    +\int_0^\infty\frac{d\om}{\pi}g\left(\frac{\om+\mu}{T}\right)
    \frac{\pa}{\pa\om}\delta(\om)
}
with
\eq{9}{ g(x) &- \frac{x}{e^x-1}-\ln\left(1-e^{-x}\right).
}
We mention that the  integrals here and in \Ref{4} are convergent due to the Boltzmann factor resp. the function $g$, \Ref{9}. Hence no regularization is necessary and one can get numerical results easily.

It is interesting to consider the behavior for high temperature. For the entropy, expanding the function $g$ for small argument,
\eq{11}{ g(x) &= -\ln(x)+1+O(x)
}
one arrives at
\eq{10}{ S   & \raisebox{-4pt}{$\sim\atop T\to\infty$}
    \sum_n\left(\ln(T)+1-\ln(-\kappa_n+\mu)\right)+
    \frac{1}{\pi}\int_0^\infty d\om\,\left(\ln(T)+1-\ln(\om+\mu)\right)
         \frac{\pa}{\pa\om}\delta(\om) .
}
In the examples considered in this paper the integral over $\om$ in \Ref{10} will be convergent.
The behavior for low $T$ depends on the details of the models and can be obtained simply by expanding the phase shift for small $\om$.

\section{\label{T3}Entropy for plasma point}
Here we consider the solutions of eq. \Ref{1} with potential and boundary condition given in \Ref{2}. We put the chemical potential $\mu=0$ since we consider $\Om>0$ only. The solutions of eq. \Ref{1} with the potential \Ref{2} reads
\eq{12}{ \Phi(x) &= \sin(\om x)\Theta(R-x)+\left(\sin(\om x)+\frac{\Om}{\om}\sin(\om R)\sin(\om(x-R))\right)\Theta(x-R),
}
as can be checked easily. This solution can be rewritten in the form
\eq{13}{ \Phi(x) &= \sin(\om x)\Theta(R-x)
        +\frac12 \left(f(\om)\, e^{-i\om x}+f(\om)^* e^{i\om x}\right)\Theta(x-R)
}
where the coefficients
\eq{14}{ f(\om) &= \frac{1}{i}\left(1+\frac{\Om}{\om}\sin(\om R)e^{i\om R}\right)
}
are the Jost functions (and its complex conjugate). As mentioned, we consider $\Om>0$ only so that there are no bound states.

\begin{figure}[t]\label{fig2}
\begin{minipage}{0.48\textwidth}
    \includegraphics[width={0.98\textwidth}]{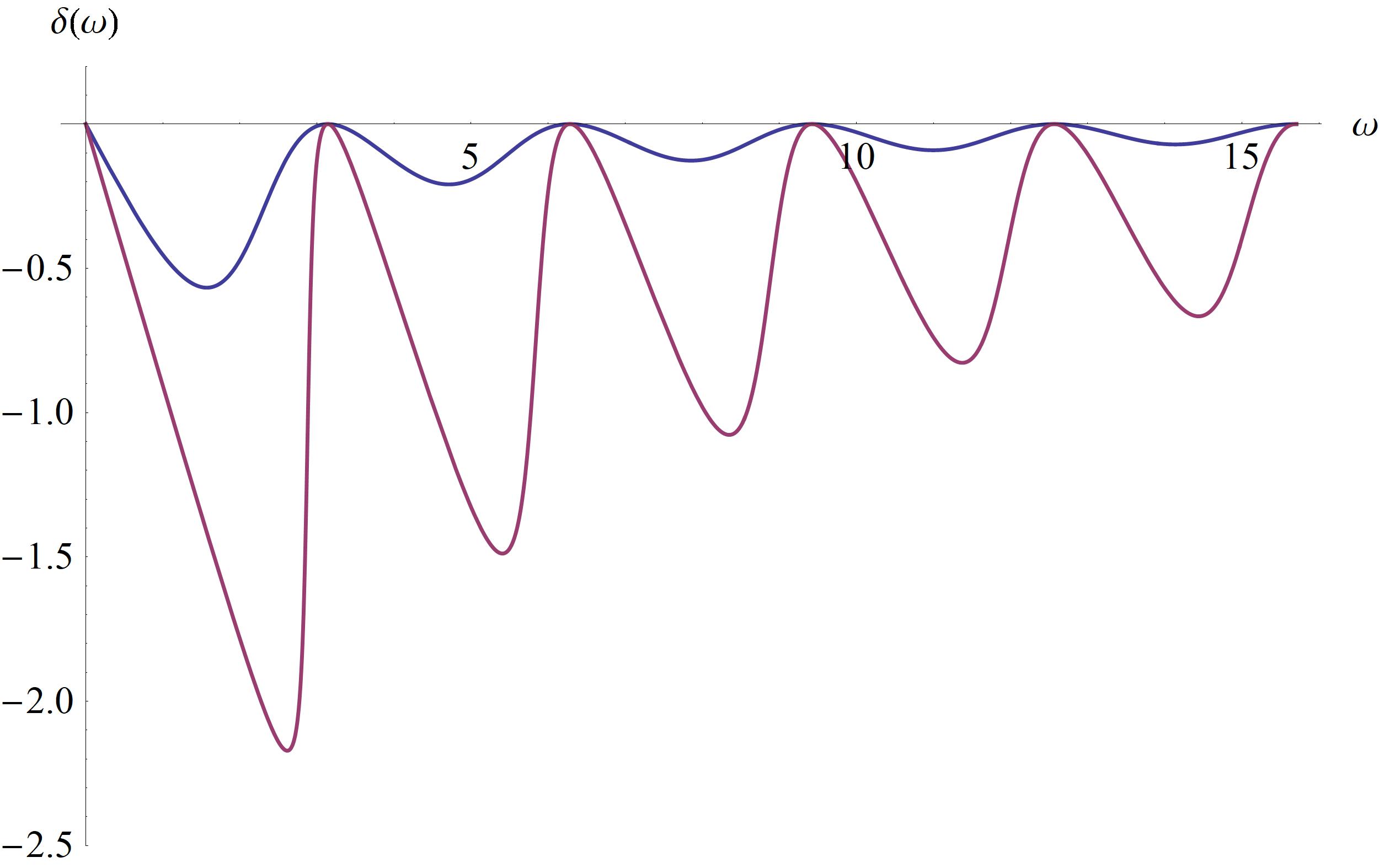}
    \caption{The phase shift \Ref{15} as function of $\om$ for $\Om R=1$ (upper curve) and
    $\Om R=10$ (lower curve).
    }
\end{minipage} \ \
\begin{minipage}{0.48\textwidth}
    \includegraphics[width={1\textwidth}]{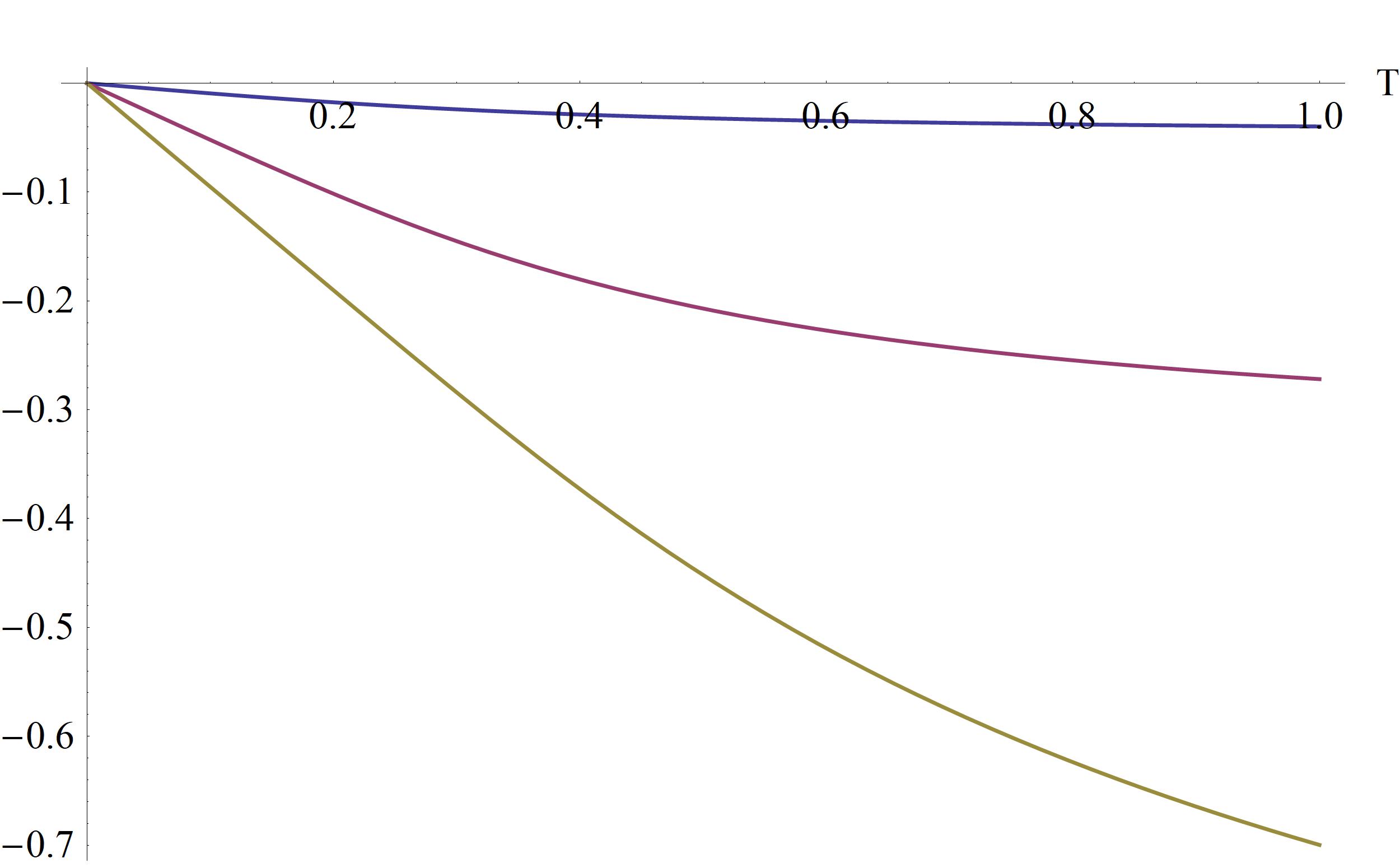}
    \caption{The entropy $S$, \Ref{8}, as function of $T$  for the plasma point for $\Om R=0.1,\ 1, \ 10$ (from top to bottom). }
\end{minipage}\ \ \
\end{figure}

The phase shift is given by
\eq{14a}{ \delta(\om) &=\frac{-1}{2i}\ln\left(\frac{f(\om)}{f(\om)^*}\right)
}
instead of  \Ref{5} since we have a problem on the half axis and inserting \Ref{14} we get
\eq{15}{ \delta(\om) &= -\frac{\pi}{2}+\arctan\left(\frac{1+\frac{\Om}{\om}\sin(\om R)\cos(\om R)}{\frac{\Om}{\om}\sin^2(\om R)}\right).
}
This coincides with $\delta_\ell(\om)$ for $\ell=0$ (s-wave) in the TE polarization in \cite{Text18-1}.

Thew asymptotic expansions for small and large argument are
\eq{16}{ \delta(\om) & \raisebox{-4pt}{$\sim\atop \om\to 0$}
        -\frac{\Om R}{1+\Om R}\, \om,
\ \ \ & \delta(\om) & \raisebox{-4pt}{$\sim\atop \om\to \infty$}
        -\Om\sin^2(\om R)\frac{1}{\om}\,,
}
and   plots are shown in Fig. 1 for several $\Om$.

Inserting the phase shift \Ref{15} into eq. \Ref{8} for the entropy allows for easy numerical evaluation. Examples are shown in Fig. 2. As can be seen, the entropy is completely negative.

The behavior of the entropy for $T\to\infty$ can be obtained inserting the phase shift \Ref{15} into \Ref{10},
\eq{17}{S &= \frac{1}{\pi}\int_0^\infty d\om\, \left(\ln(T)+1-\ln(\om)\right)
\frac{\pa}{\pa\om}\delta(\om)+O\left(\frac{1}{T}\right),
\nn\\   &= -\frac12\ln(1+\Om R)+O\left(\frac{1}{T}\right),
}
where we used \Ref{16} for the integral over a derivative. The integral involving $\ln(\om)$ could be done explicitly. Its sign can be seen after integrating by parts,
\eq{17a}{ S&=\frac{1}{\pi}\int_0^\infty\frac{d\om}{\om}\, \delta(\om)+O\left(\frac{1}{T}\right),
}
and looking on Fig. 1.

\section{\label{T4}Entropy for single delta potential}
Here we consider the model given by the equation \Ref{1} together with the potential \Ref{3}. The phase shifts \Ref{5} are well known and can be written in the form
\eq{18}{ \delta(\om) &= \left\{\begin{array}{cc}
    -\frac{\pi}{2}+\arctan\left(\frac{\om}{\al}\right), & (\al>0),
    \\[4pt] \frac{\pi}{2}+\arctan\left(\frac{\om}{\al}\right), & (\al<0),
    \end{array}\right.
}
ensuring vanishing phase shifts for $\om\to\infty$. These are shown in Fig. 3.

\begin{figure}[h]\label{fig4}
\begin{minipage}{0.48\textwidth}
    \includegraphics[width={0.98\textwidth}]{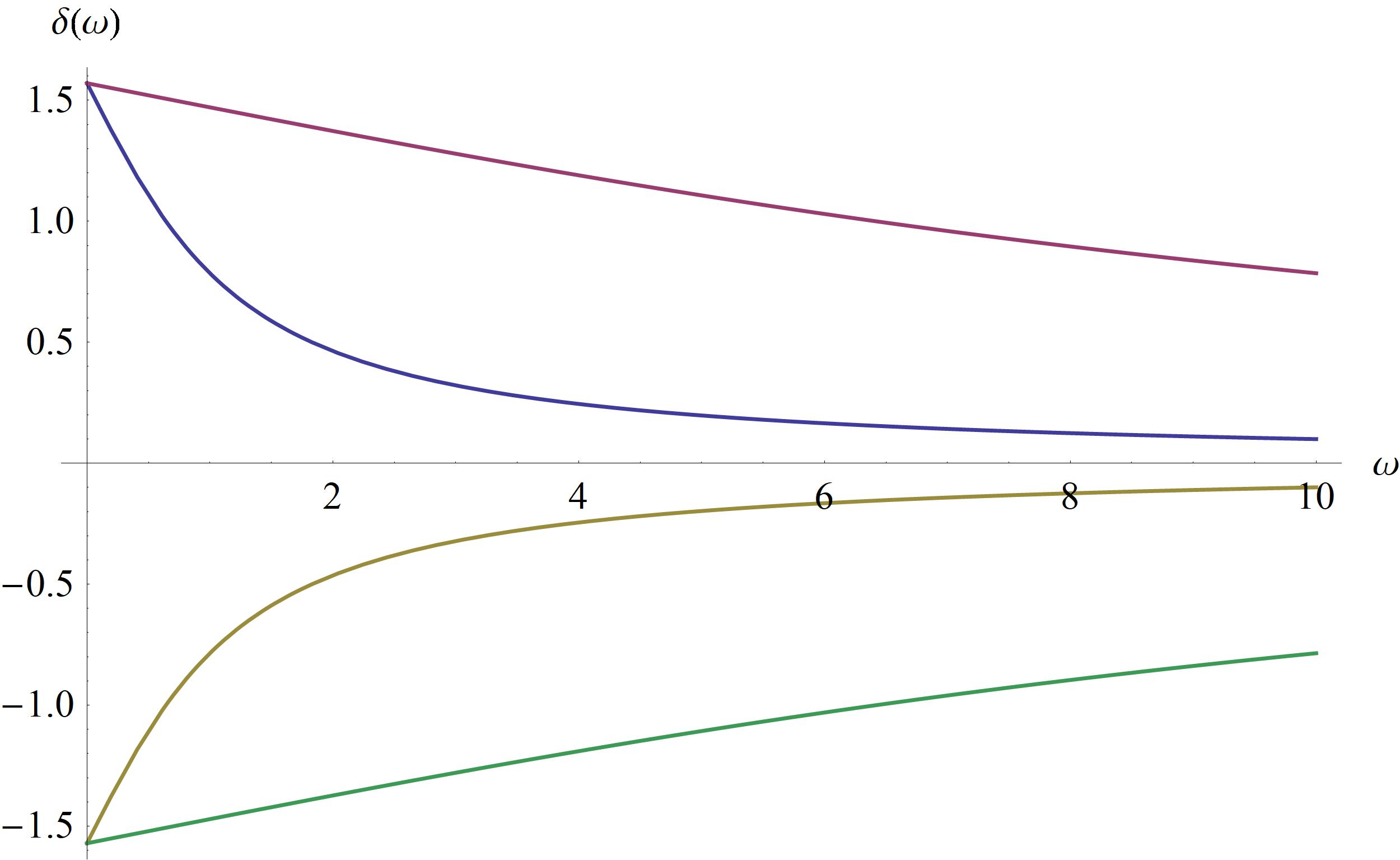}
    \caption{The phase shift \Ref{18} as function of $\om$ for $\al =-10,\ -1,\ 1\ 10$ (from top to bottom).
    }
\end{minipage} \ \
\begin{minipage}{0.48\textwidth}
    \includegraphics[width={1\textwidth}]{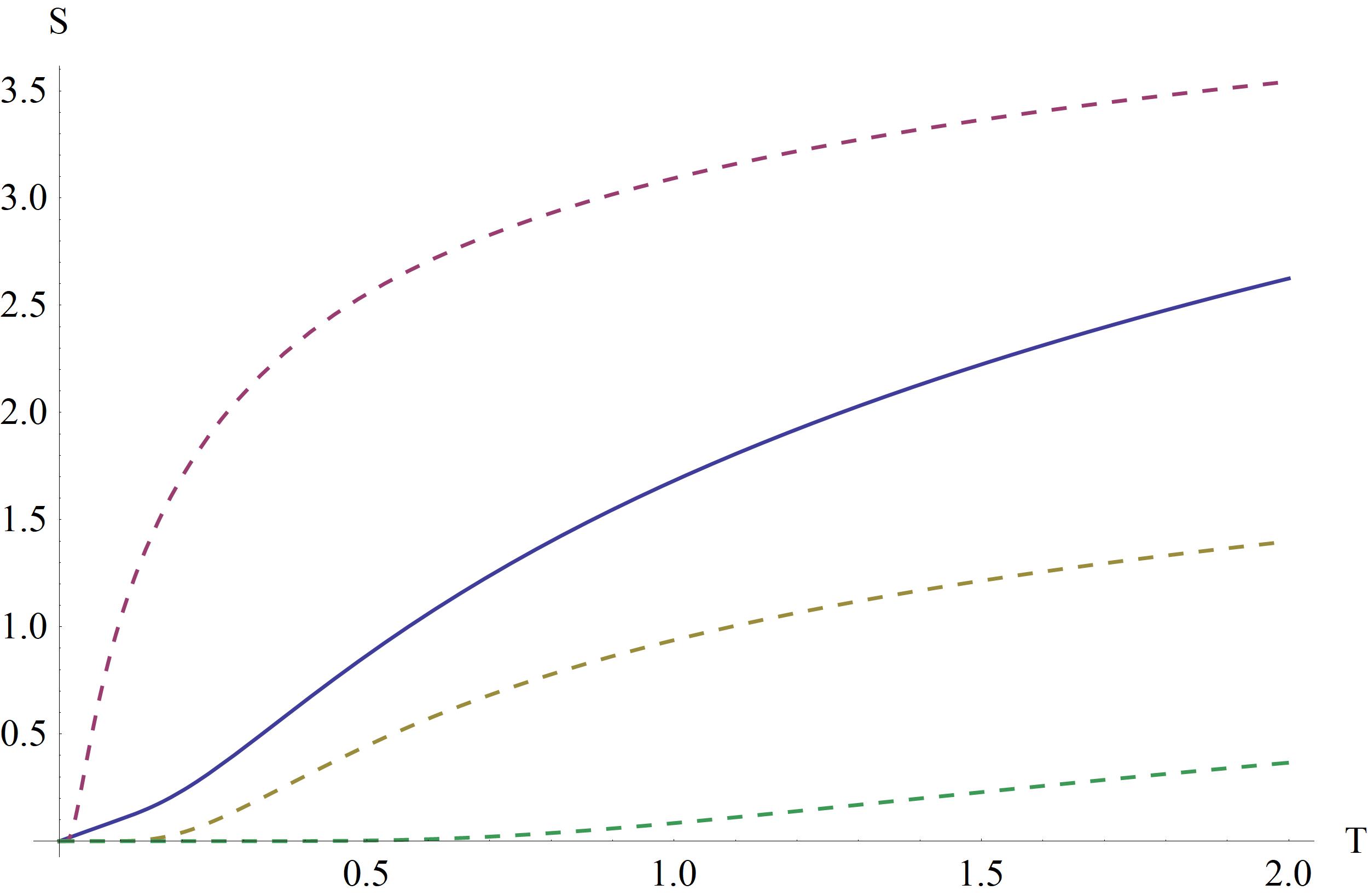}
    \caption{The entropy $S$, \Ref{8}, as function of $T$  for the delta function potential \Ref{3}   for $\al =1$ (solid line) and for $\al =-1$ with $\mu=1,\ 1.1,\ 2$ (dashed lines from top to bottom). }
\end{minipage}\ \ \
\end{figure}

Using these phase shifts in \Ref{8}, the entropy can easily be evaluated numerically. Results are shown in Fig. 4. For $\al<0$ we have a bound state with $\kappa=\al $ and  included chemical potential $\mu$. As seen, in all cases the entropy is positive.

The behavior for $T\to\infty$ can be obtained similar to \Ref{17}. However, in this case we have a contribution to the logarithmic part from $\delta(\infty)-\delta(0)=\frac{\pi}{2}{\rm sign}(\al)$ and because of
$\int_0^\infty dp\,\ln(\om) \delta'(\om)=0$ with $\delta(\om)$ from \Ref{18} no constant contribution. For $\al<0$ we have to account for the bound state contribution in \Ref{10}. We find
\eq{19}{ S &= \frac12 (\ln(T)+1) +O\left(\frac{1}{T}\right) ,
}
which is in agreement with Fig. 4.

\section{\label{T5}Applying Levinson's theorem}
Levinson's theorem is a statement on the relation between the number of bound states and the difference
\eq{20}{ \Delta\delta &=\delta(0)-\delta(\infty)
}
between the phase shifts at zero and at infinite energies. Originally it was established in \cite{levi49-25-1} for three-dimensional scattering in quantum mechanics, later derived for the one-dimensional case in \cite{jack75-12-1643} (for an almost complete bibliography see \cite{dong00-39-469}). In the one-dimensional case it states
\eq{21}{ \Delta\delta_+ +\frac{\pi}{2} &= \pi N_+,& \Delta\delta_- &=\pi N_-, &\ \ \ \mbox{for the noncritical case},
\nn \\   \Delta\delta_+    &= \pi N_+, & \Delta\delta_- -\frac{\Pi}{2} &=\pi N_-, &\ \ \ \mbox{for  noncritical case (half bound state)}.
}
Here, $N_+$ and $N_-$  are the number of bound states with even  and odd parity and $\Delta\delta_+$ and $\Delta\delta_-$ are the corresponding differences of phase shifts of the scattering states with the same parity as stated in  \cite{dong00-39-469}.

Now we consider the entropy S for $T\to\infty$ as given by eq. \Ref{10} and consider only the logarithmic contribution. We get
\eq{22}{ S &= \ln(T)\left(\sum_n 1+\frac{1}{\pi}\int_0^\infty d\om\,\frac{\pa}{\pa\om}\delta(\om)\right)+O(1)
\nn\\ &= \ln(T)\left(N-\frac{1}{\pi}\delta(0)\right)+O(1)
}
where we assumed $\delta(\infty)=0$ and $N$ is the number of bound states. From the theorem we get
\eq{23}{ N-\frac{1}{\pi} \delta(0) &=
\left\{ \begin{array}{cc}
\left\{\begin{array}{cc}\frac12 &\mbox{even}\\0 &\mbox{odd}\end{array}\right\}
& \mbox{non critical case,}
\\[8pt]
\left\{\begin{array}{cc}0 &\mbox{even}\\-\frac12 &\mbox{odd}\end{array}\right\}
& \mbox{  critical case.}
\end{array} \right.
}
Finally we account for the fact that there is always an even state before an odd appears. This way, for $T\to\infty$ the entropy can never have a negative logarithmic behavior for all configurations to which Levinson's theorem can be applied.

We mention that for the examples considered in Sections 3 and 4 the theorem is satisfied (for the spherical case it states $\Delta\delta_\ell=\pi N_\ell$ for a state with orbital momentum $\ell$, see, e.g., \cite{tayl72b}, p. 227).
\section{\label{T6}Conclusions}
We investigated the entropy for  simple one-dimensional configurations and found both, negative and positive values. We mention that in the one-dimensional case there is no need for subtracting terms growing too fast at $T\to\infty$ - a procedure which is still under discussion. We mention that we did not consider a Casimir effect like situation, where only the separation dependent part of the free energy is considered, but calculated the complete entropy of the considered systems. Having in mind the results of \cite{Text18-1} and \cite{Text18-2} together with the above ones, we conclude that the occurrence of negative entropies is rather the rule than the exception.

\end{document}